\documentclass[12pt]{amsart}
\usepackage{amsmath}
\usepackage{amsthm}
\usepackage{amsfonts}
\usepackage{graphicx}
\newenvironment{nouppercase}{
  
  \renewcommand{\uppercasenonmath}[1]{}}{}

\begin{document}

\title
{Commuting signs of infinity}
\author{Jens Hoppe}
\address{Braunschweig University, Germany}
\email{jens.r.hoppe@gmail.com}

\begin{abstract}
Discrete minimal surface algebras and
Yang Mills algebras may be related to (generalized) Kac Moody algebras,
just as Membrane (matrix) models
and the IKKT model - including a novel construction
technique for minimal surfaces.
\end{abstract}

\begin{nouppercase}
\maketitle
\end{nouppercase}
\thispagestyle{empty}
\noindent
In this note I would like to mention some aspects
of two kinds of double commutator equations\footnote{the bilinear antisymmetric bracket, assumed to satisfy the Jacobi identity,
is not assumed to necessarily come from an underlying associative multiplication,
i.e. could also be a Poisson--bracket; repeated indices are summed over, unless
stated otherwise; the distinction between upper and lower indices could of course
also be made in (2), and the distinction between (1) and (2) is equally ``pragmatical''}
\begin{equation}\label{eq1} 
\big[ [X^{\mu}, X^{\nu}], X_{\nu} \big] = 0
\end{equation}
\begin{equation}\label{eq2} 
\big[ [M_i, M_j], M_j \big] = \mu_i M_i.
\end{equation}
(\ref{eq1}) appears e.g. in \cite{JH96}\cite{IKKT96}\cite{3}\cite{CD02}\cite{ACH13}\cite{AHK19}\cite{S20}
(and references therein), describing non--commutative minimal surfaces, resp. a quantization of string theory
(related to the Schild action \cite{S77}; stronger conditions, including $\big[ [X^{\mu}, X^{\nu}], X_{\rho} \big]=0$,
implying basic uncertainty relations for a non--commutative
space--time, appear in \cite{DFR95}), resp. covariant derivatives of Yang-Mills 
connections (see e.g. \cite{12} \cite{HS8});  (\ref{eq2}) e.g. in \cite{JH82}\cite{12}\cite{B89}\cite{JH94}\cite{JH97}
\cite{AHT3/4}\cite{AH9}\cite{KKLN21} (and references therein).
\cite{B89} implies that the maximal compact subalgebra of simply laced Kac Moody algebras (for GIM algebras, see \cite{S84/86};
note that \cite{AH9} contains observations and ideas that may also be relevant for the general infinite dimensional case) 
is isomorphic 
to the quotient of a free Lie algebra generated by $Y_1 \ldots Y_n$ subject to the relations 
\begin{equation}\label{eq3} 
\big[ [Y_i, Y_j], Y_j \big] = \pm Y_i \qquad \text{({\it no} sum)}
\end{equation}
if the $(ij)$ entry of the generalized Cartan matrix $A$ is non--zero, while 
\begin{equation}\label{eq4} 
[Y_i, Y_j] = 0 \quad \text{if} \quad A_{ij} = 0
\end{equation}
(the sign in (\ref{eq3}) is in principle fixed by the Cartan involution and reality properties; it would be interesting to see whether {\it vanishing} `signs' could arise after summing over $j$, thus relating (\ref{eq3}) also to (\ref{eq1})). The relation to (\ref{eq2}) is apparent, as for each generalized Cartan matrix $\mu_i$ simply results from the number of non--zero  elements $A_{j\neq i}$ in the $i$--th column resp. row (which e.g. for the affine Kac Moody algebra $\tilde{A}_l$ would be $2$, independent of $i$). Note that in the relation of (\ref{eq2}) with
the membrane (matrix) model in $D$--dimensional space--time \cite{JH82}\cite{JH13}, with classical equations of motion
\begin{equation}\label{eq5} 
\ddot{X}_i = -\sum^d_{j=1}\big[ [X_i, X_j], X_j\big],
\end{equation}
the $X_i$ being $d=D-2$ time--dependent traceless hermitean $N \times N$ matrices, one would want the non-zero $\mu_i \, (>0)$ to appear in pairs -- because of the Ansatz (cp. e.g. \cite{AH9})  
\begin{equation}\label{eq6} 
X_i(t) = \big( e^{J(t-t_0)}\big)_{ij} M_j
\end{equation}
with $J^T = -J$ real, $J^2$ diagonal -- and the hermitean $M_i$ to satisfy (\ref{eq2}), as well as
\begin{equation}\label{eq7} 
J_{ij}[M_i, M_j] = 0
\end{equation}
(in order to satisfy the $SU(N)$--`Gauss--constraint' $\sum[X_i, \dot{X}_i] = 0$, which is the discrete analogue of the residual invariance under area--preserving diffeomorphisms \cite{JH82}, hence has to be satisfied for (\ref{eq5}) to include membranes as $N\rightarrow \infty$). One\footnote{in the context of membrane (matrix) solutions `practised' -- though I always considered it as somewhat unnatural.}
way to satisfy the constraint would be to choose half of the $M$'s to be identically zero, while
in view of the Berman--construction, (\ref{eq3})+(\ref{eq4}), one could simply pair each node with one to which it is not connected; or consider (\ref{eq7}) an analogue of the sum--condition in (6.3) of \cite{AHK19}. Of particular interest would be to identify the maximal compact subalgebra of $E_{10}$ ($E_9$) in (\ref{eq1}) ((\ref{eq2})/(\ref{eq5})).\\[0.15cm]
In the simplest example, $\tilde{A}_l$, one gets
\begin{equation}\label{eq8} 
\sum_j \big[ [Y_i, Y_j], Y_j \big] = 2Y_i, 
\end{equation}
each simple root having 2 neighbours; and
the simplest finite--dimensional generalized spin representation of (\ref{eq3})+(\ref{eq4}), hence (\ref{eq8})/(\ref{eq2}), in this case is (apparently \cite{12} first noticed by A.Kent; though the connection with infinite dimensional Lie--algebras was not
realized at that time)
\begin{equation}\label{eq9} 
\begin{split}
M_k & := Y_k := \frac{i}{2}\gamma^k \gamma^{k+1} = \frac{i}{2}\gamma^{k\,k+1}\\
M_k & = M^{\dagger}_k, \quad M^2_k = \frac{1}{4}, \qquad \qquad k= 1 \ldots K \; (K+1 \equiv 1)
\end{split}
\end{equation}
where the $\gamma^k$ are traceless anti--commuting hermitean (Clifford) matrices squaring to $\mathbf{1}$.
In the context of the $d=9$ rotating membrane solutions example one could e.g. take $K=8$, $\gamma_1 = \sigma_1 \times 1 \times 1 \times 1$, $\gamma_2 = \sigma_2 \times 1 \times 1 \times 1$, $\gamma_3 = \sigma_3 \times \sigma_1 \times 1 \times 1$, \ldots, $\gamma_8 = \sigma_3 \times \sigma_3 \times \sigma_3 \times \sigma_2$, $N=16$, resp. $8$ (note that the Clifford--solutions of (\ref{eq2}) in \cite{AH9} naively would need the doubling mechanism, i.e. $K=4$, $N=4$).\\
For the affine Kac--Moody algebra $\tilde{D}_l$ one would naturally get solutions of (\ref{eq2}) where $\mu =1$ has multiplicity 4, $\mu = 3$ multiplicity 2, and $\mu = 2$ multiplicity $l-5$.\\
While in the physics context the most important aspect of realizing (\ref{eq3})+(\ref{eq4}) is that it signals  potential infinite symmetries  for (\ref{eq2}) (resp.(\ref{eq1}); note the `reconstruction algebra' \cite{JH11}\cite{JH13}), including a possible relation to the area--preserving diffeomorphism algebra for relativistic extended objects \cite{JH82} \cite{JH13}, there is another, equally interesting, aspect: (\ref{eq3}) can (and does) describe discrete minimal surfaces (hence the name DMSA in \cite{AH9}) embedded in spheres (once in each connected component the $\mu_i$ are equal and the constraint $\sum M^2_i \sim \mathbf{1}$ is added); hence
it is natural to conjecture that the generalized higher spin representations of (\ref{eq3})+(\ref{eq4}) (cp. \cite{KKLN21}, and references therein) include series of finite dimensional representations (of increasing dimension) that for $N\rightarrow \infty$ converge to (new) minimal surfaces in spheres.

\vspace{1.5cm}
\noindent
\textbf{Acknowledgement.} 
The above note originated when trying to understand 
the representation theory of area preserving diffeomorphisms.
While I could not yet make much progress in that direction,
I am grateful to J. Arnlind, V. Bach, J. Eggers, M. Floratos, J. Fr\"ohlich, M. Hanada, M. Hynek,
G. Ishiki, N. Iyudu,
 V. Kac, A. Kleinschmidt,
B. Khesin,  P. Michor, 
D. O'Connor,
T. Ratiu,
R. Suter 
and T. Turgut for discussions
and correspondence.


\begin{thebibliography}{11111}
\setlength{\baselineskip}{0.9\baselineskip}
\bibitem[1]{JH96} J.Hoppe, {\it On M--Algebras, the Quantization of Nambu--Mechanics, and Volume Preserving Diffeomorphisms}, arXiv:hep-th/9602020 
\bibitem[2]{IKKT96} N.Ishibashi, H.Kawai, Y.Kitazawa, A.Tsuchiya, {\it A Large-N Reduced Model as Superstring}, arXiv:hep-th/9612115
\bibitem[3]{3} L.Cornalba, W.Taylor IV, {\it Holomorphic curves from matrices}, Nucl.Phys.B 536 (1999) 513
\bibitem[4]{CD02} A.Connes, M.Dubois-Violette, {\it Yang-Mills algebra}, arXiv:math/0206205
\bibitem[5]{ACH13}  J.Arnlind, J.Hoppe, {\it The world as quantized minimal surfaces}, arXiv:1211. 1202,
					J.Arnlind, J.Choe, J.Hoppe, {\it Noncommutative Minimal Surfaces}, arXiv: 1301.0757
\bibitem[6]{AHK19} J.Arnlind, J.Hoppe, M.Kontsevich, {\it Quantum Minimal Surfaces}, arXiv: 1903.10792
\bibitem[7]{S20} H.Steinacker, {\it Quantum (Matrix) Geometry and Quasi-Coherent States}, arXiv:2009.03400
\bibitem[8]{S77}  A.Schild, {\it Classical null strings}, Phys.Rev.D16 (1977) 1722
\bibitem[9]{DFR95} S.Doplicher, K.Fredenhagen, J.E.Roberts, {\it The Quantum Structure of Spacetime at the Planck Scale and Quantum Fields}, CMP 172 (1995) 187
\bibitem[10]{HS8} A.Connes, M.Dubois-Violette {\it Yang-Mills and some related algebras}, arXiv:math-ph/0411062,
                  R.Berger, M.Dubois-Violette {\it Inhomogeneous Yang-Mills algebras}, arXiv:math/0511521,
                  E.Herscovich, A.Solotar, {\it Representation theory of Yang-Mills algebras}, arXiv:0807.3974, 
\bibitem[11]{JH82} J.Hoppe, {\it Quantum theory of a massless relativistic surface}, Ph.D. thesis, MIT 1982 http://dspace.mit.edu/handle/1721.1/15717,
                   J.Hoppe, H.Nicolai, {\it Relativistic Minimal Surfaces}, Phys.Lett.B 196 (1987) 451
\bibitem[12]{12}  E.Corrigan, P.R.Wainwright, S.M.J.Wilson {\it Some Comments on the non self-dual Nahm equations}, Comm.Math.Phys. 98 (1985) 259,  
E.Corrigan, {\it Some Comments on a Cubic Algebra}, Proceedings of the 1985 Snri Winter School `Geometry and Physics'
published in: Rendiconti Circolo Matematico di Palermo, Serie II, Supplemento \#9 (1985) 43
\bibitem[13]{B89} S.Berman, {\it On generators and relations for certain involutory subalgebras of Kac-Moody Lie algebras}, Comm.Algebra 17 \#12 (1989) 3165
\bibitem[14]{JH94} J.Hoppe, {\it Some Classical Solutions of Membrane Equations in 4 Space-Time Dimensions}, arXiv:hep-th/9402112
\bibitem[15]{JH97} J.Hoppe, {\it Some Classical Solutions of Membrane Matrix Model Equations}, arXiv:hep-th/9702169
\bibitem[16]{AHT3/4} J.Arnlind, 
J.Hoppe arXive:hep-th/0312062/0312166,
J.Hoppe, S.Theisen arXive:hep-th/0405170,
J.Arnlind, J.Hoppe, S.Theisen, {\it Spinning Membranes}, Phys.Lett.B 599 (2004) 118
\bibitem[17]{AH9} J.Arnlind, J.Hoppe, {\it Discrete Minimal Surface Algebras}, arXive:0903.5237 
\bibitem[18]{KKLN21} A.Kleinschmidt, R.K\"ohl, R.Lautenbacher, H.Nicolai, {\it Representations of Involutory Subalgebras of Affine Kac-Moody Algebras}, arXiv:2102.00870
\bibitem[19]{S84/86} P.Slodowy, {\it Singularit\"aten: Kac-Moody-Liealgebren, assoziierte Gruppen und Verallgemeinerungen}, Habilitationsschrift Universit\"at Bonn 1984
\bibitem[20]{JH11} J.Hoppe {\it Fundamental structures of M(brane) theory}, Phys.Lett.B695 (2011) 384
\bibitem[21]{JH13} J.Hoppe, {\it Relativistic Membranes}, J.Phys.A 46 (2013) 023001

\end{thebibliography}
\end{document}